\documentclass[aps,prd,reprint,superscriptaddress,preprintnumbers,nofootinbib]{revtex4-1}
\usepackage{amsmath}
\usepackage{amsfonts} 
\usepackage{amssymb}
\usepackage{graphicx}
\usepackage{hyperref}
\usepackage{tensor}
\usepackage{siunitx}
\usepackage{float}
\newcommand{\be}{\begin{equation}}
\newcommand{\ee}{\end{equation}}

\begin{document}

\preprint{}
\preprint{}

\title{\Large Fermions and the Swampland}

\author{\large Eran Palti}
\affiliation{Max-Planck-Institut f\"ur Physik (Werner-Heisenberg-Institut),
             F\"ohringer Ring 6,
             80805, M\"unchen, Germany}
             \affiliation{Department of Physics, Ben-Gurion University of the Negev, Beer-Sheva 84105, Israel}

\begin{abstract}
\vspace{0.5cm}
In this note we consider whether there could be Swampland constraints associated to the presence of fermions in the theory.  We propose that any fermion must couple to an infinite tower of states, and that the mass scale of this tower, in Planck units, is set by the strength of the Yukawa coupling to the tower. This is a type of fermionic version of the (magnetic) Weak Gravity Conjecture. We also find that supersymmetry plays a natural part in this fermionic realisation, which motivates a further proposal that supersymmetry can only be broken below the scale set by this Yukawa coupling. We perform some preliminary checks in string theory of these ideas.
\vspace{1cm}
\end{abstract}

\maketitle

\section{Introduction}
\label{sec:int}

\begin{figure}
\centering
 \includegraphics[width=0.45\textwidth]{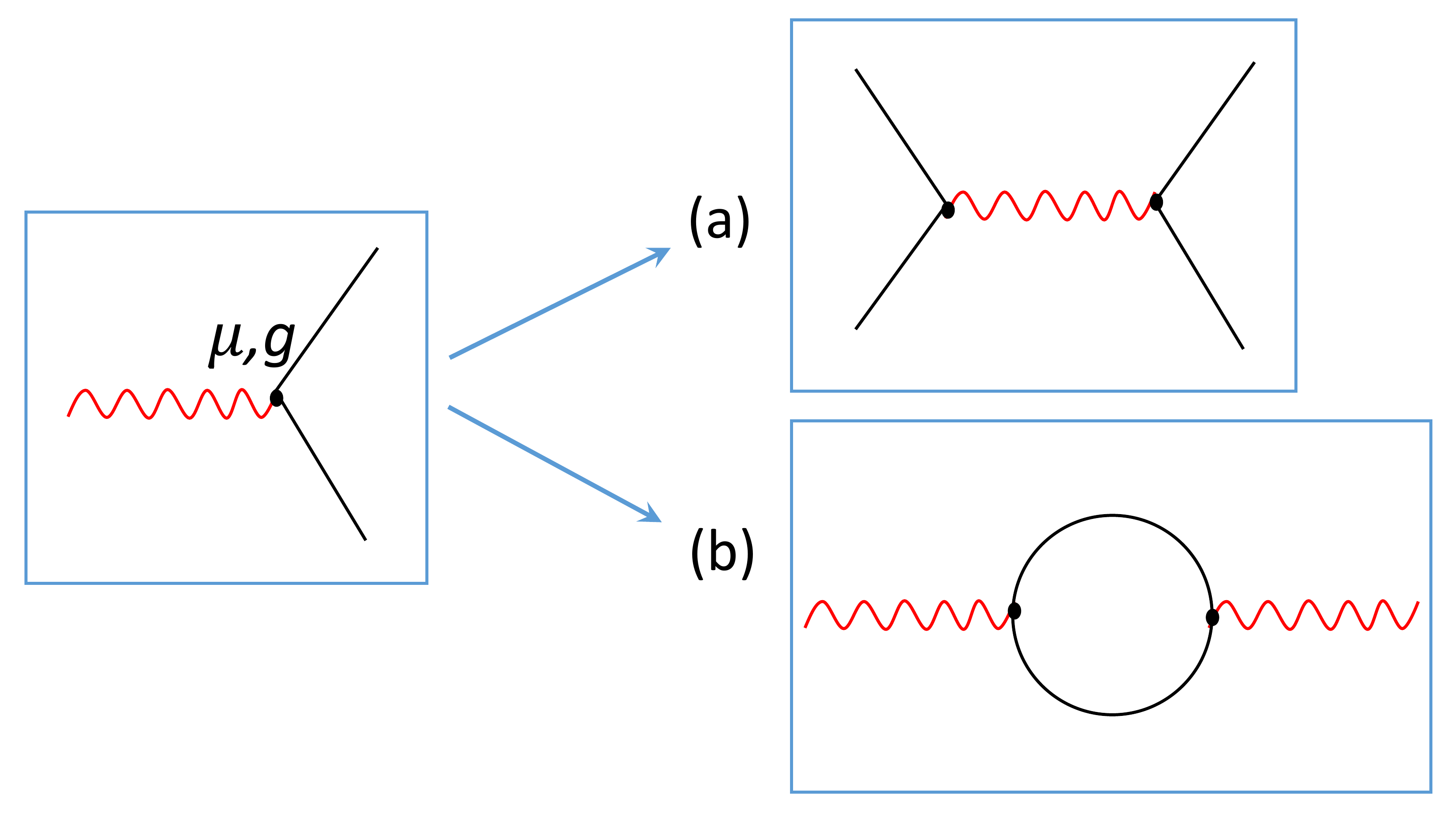}
\caption{Figure showing two Feynman diagrams in which the cubic couplings of (\ref{mwgc}) and (\ref{sdc}) appear. Case (a) is associated to forces and possibly to black hole discharge. Case (b) is instead associated to running of the gauge coupling and the emergence proposal.}
\label{fig:fey}
\end{figure}

The Swampland program aims to establish constraints on effective theories coming from a consistent ultraviolet completion to quantum gravity (see \cite{Vafa:2005ui} for initial work and \cite{Brennan:2017rbf,Palti:2019pca} for reviews). Two important such proposed constraints are the Weak Gravity Conjecture (WGC) \cite{ArkaniHamed:2006dz} and the Swampland Distance Conjecture (SDC) \cite{Ooguri:2006in} (see also \cite{Baume:2016psm,Klaewer:2016kiy} for a refined version). The magnetic WGC, or more precisely extensions of it \cite{Heidenreich:2015nta,Klaewer:2016kiy,Montero:2016tif,Heidenreich:2016aqi,Andriolo:2018lvp,Grimm:2018ohb}, proposes that given a gauge field with gauge coupling $g$, there must be an infinite tower of states with a mass scale $m^g_{\infty}$ which satisfies
\be
m^g_{\infty} \sim g M_p \;.
\label{mwgc}
\ee
The SDC proposes that given a canonically normalised scalar field $\phi$ there exists an infinite tower of states with a mass scale $m^{\mu}_{\infty}$ depending exponentially on $\phi$ (for large $\phi > M_p$), so schematically $m^{\mu}_{\infty} \sim e^{-\alpha \phi} M_p$ for some $\alpha \sim {\cal O}(1) M_p^{-1}$. We can rewrite this statement in a suggestive way, by defining $\mu \equiv \left|\partial_{\phi} m^{\mu}_{\infty} \right|$, as
\be
m^{\mu}_{\infty} \sim \mu M_p \;.
\label{sdc}
\ee
There is a clear similarity (pointed out in \cite{Palti:2017elp}) between (\ref{mwgc}) and (\ref{sdc}) once we note that both $\mu$ and $g$ are the values of the cubic couplings of the associated field to the massive tower of states. These cubic couplings appear in two types of diagrams, in the exchange force diagram and in the 1-loop self-energy diagrams. This is illustrated in figure \ref{fig:fey}.

The proposals (\ref{mwgc}) and (\ref{sdc}) seem to be respected in all known string theory examples. But what could be the physics underlying them? One reasoning, as in the original proposal \cite{ArkaniHamed:2006dz}, is that (\ref{mwgc}) could be related to black holes and monopoles, or to a requirement that gravity should always be the weakest self-force \cite{ArkaniHamed:2006dz,Palti:2017elp,Heidenreich:2019zkl}. In that case the conditions on the couplings would be due to their appearance in exchange diagrams like (a) in figure \ref{fig:fey} (see, for example, \cite{Lust:2017wrl,Lee:2018spm,Gonzalo:2019gjp,Freivogel:2019mtr,DallAgata:2020ino,Gendler:2020dfp,Benakli:2020pkm} for studies of relevant interaction strengths).

A different (though possibly related) proposal for the physics underlying (\ref{mwgc}) and (\ref{sdc}) comes from considering wavefunction renormalization. This is the so-called emergence proposal \cite{Harlow:2015lma,Heidenreich:2017sim,Grimm:2018ohb,Heidenreich:2018kpg,Palti:2019pca}. We can rewrite (\ref{mwgc}) suggestively as \cite{Heidenreich:2017sim,Grimm:2018ohb}
\be
\frac{1}{g^2} \sim \sum_{n=1}^N  q^2_n \;,
\label{sumgn}
\ee
where $q_n = n$ is the (increasing) charge of a state in the tower, and the sum is up to $N$ states in the tower with 
\be
N^3 \sim \left(\frac{M_p}{m^g_{\infty}}\right)^2 \;.
\label{Nrel}
\ee
The rewriting in (\ref{sumgn}) is such that the right hand side corresponds to the running of the gauge coupling, due to diagrams of type (b) in figure \ref{fig:fey}, from integrating out $N$ states in the tower of charged states. Setting this approximately equal to the left hand side is the statement that the gauge coupling receives an order one contribution in the infrared from integrating out these states. 

Note that within the emergence proposal (\ref{sumgn}) is taken to mean that the gauge coupling is generated by the tower of states such that any ultraviolet contribution at the top of the tower (or more precisely after $N$ states) is restricted to be order one or vanishing.\footnote{This is actually a self-consistency requirement on the proposal in order to keep the tower contribution to the kinetic term of order one at any scale up to the top of the tower.} This can equivalently be understood as the statement that the gauge field becomes strongly coupled at the scale 
\be
\Lambda_s \sim N m^g_{\infty} \sim \frac{M_p}{\sqrt{N}} \;,
\label{spsc}
\ee
which is the so-called Species scale \cite{PhysRevD.34.373,Bekenstein:1993dz,Jacobson:1994iw,Bekenstein:2000sw,Veneziano:2001ah,Dvali:2001gx,ArkaniHamed:2005yv,Distler:2005hi,Dimopoulos:2005ac,Dvali:2007hz,Dvali:2009ks,Palti:2019pca,Dvali:2020wqi}. This is the natural scale at which gravity becomes strongly coupled.

The same analysis applies to the case of scalar fields as in (\ref{sdc}) \cite{Grimm:2018ohb,Heidenreich:2018kpg}. Therefore, both (\ref{mwgc}) and (\ref{sdc}) are explained by a requirement that their kinetic terms in the infrared are of order the contribution to them from integrating out the tower of states at one-loop. This can be motivated as a type of unification at a single scale, or as evidence that the dynamics of the fields are emergent in the infrared from integrating out the tower of states (so they are not dynamical fields at the species scale). 

There are a number of open questions about this analysis. The most prominent being that there is no known microscopic theory valid at the species scale in which the integrating out can be explicitly performed. Nonetheless, at least in toy-models of emergent gauge fields which are calculable, the kinetic term for the gauge fields does behave like the naive integrating out of massive charged states (see \cite{Harlow:2015lma,Palti:2019pca} for a discussion of this in the present context). We will not discuss these issues in this note, only study the application of the emergence proposal assuming it has some validity.

\section{Distinguishing couplings from expectation values}
\label{sec:dis}

The constraint (\ref{sdc}) is a way to state the distance conjecture for $\phi > M_p$. However, the emergence proposal states that it should hold by itself, which is a more general statement than the distance conjecture, and it is important to make this distinction. The constraint (\ref{sdc}) is proposed to hold as behaviour associated to having $\mu \ll 1$, and should be most accurate in a parametric limit $\mu \rightarrow 0$. But this is unrelated, in general, to the expectation value of $\phi$, so should hold even if $\phi$ has a vanishing expectation value. In other words, the limits $\phi \rightarrow \infty$ and $\mu \rightarrow 0$ are in general completely independent.\footnote{The constraint (\ref{sdc}) as independent of the distance conjecture or emergence proposal was first proposed in \cite{Palti:2017elp,Lust:2017wrl}. Its most natural name would be the magnetic Scalar Weak Gravity Conjecture, within the context of \cite{Palti:2017elp,Grimm:2018ohb}.} The distance conjecture is only the sub-case of (\ref{sdc}) where $\mu$ and $\phi$ are correlated such that $\phi \rightarrow \infty \implies \mu \rightarrow 0$. Of course, since string theory has no free parameters, $\mu$ is still set by an expectation value of some scalar field, and $\mu \rightarrow 0$ is expected to correspond to some infinite distance limit in field space, but not necessarily the field space of $\phi$.

Given the independence of (\ref{sdc}) from the distance conjecture, there is an important point to clarify about its meaning. The coupling $\mu$ is taken to be the coupling to an infinite tower of states, not to any particular state. So it is perfectly allowed to have $\mu=0$ for the coupling of a scalar to any particular state. However, for any scalar, there should exist an infinite tower of states with an associated coupling $\mu$ such that the mass scale of the tower respects (\ref{sdc}). In this sense it is effectively a statement about the kinetic terms of the scalar, which affect the coupling to all states.

There are not many studies in string theory of (\ref{sdc}) for cases where the coupling $\mu$ is independent of the scalar field expectation value. Such checks would actually be very important as test of the emergence proposal, and in some sense are prerequisite to the analysis for fermions presented in this note. We have performed a small preliminary analysis of some string theory tests and discuss them in appendix \ref{app:sttest}.

\section{Applying to Fermions}
\label{sec:fer}

In the emergence motivation, the couplings are constrained because they appear in diagrams of type (b) in figure \ref{fig:fey}. An important difference between diagrams of type (a) and (b) is that diagrams of type (b) can be considered with the gauge field replaced by a fermion, while there is no fermionic self-force possible in type (a) diagrams. 
%Secondly, the momenta of the particles in the tower should be vanishing for a self-force type motivation from diagram (a). While in diagram (b) their momenta are non-vanishing. This is an important difference if we consider cubic couplings which involve derivatives.

The emergence proposal can therefore be readily applied to fermions. This would lead to a natural Swampland constraint: given a fermion (either spin-$\frac12$ or spin-$\frac32$) in the effective theory, there must exist an infinite tower of states such that the kinetic term of the fermion receives an order one contribution from integrating the tower out. 

First we consider the generic case, where $\chi$ is a spin-$\frac12$ fermion which has Yukawa couplings to the infinite tower of states. Because of its fermionic nature we cannot write a coupling completely analogous to the gauge field and scalar field, where it is the same state running in the loop. Rather, the loop must involve a fermion and a scalar. Therefore, the most natural coupling involves a tower where at the $n^{\mathrm{th}}$ level there is a scalar $h_n$ and a fermion $\psi_n$, which couple as 
\be
Y_n h_n \chi \psi_n \;.
\label{Yukco}
\ee
We let the coupling strength increase up the tower proportional to the level in the tower
\be
Y_n \sim Y n \;.
\label{Yukn}
\ee
For the contribution from integrating out the tower to be of order one the mass scale of the tower $m_{\infty}^Y$ is related to this coupling as 
\be
m_{\infty}^Y \sim Y M_p \;.
\label{fwgc}
\ee
The statement (\ref{fwgc}) is then the spin-$\frac12$ analogue of the spin-1 expression (\ref{mwgc}) and the spin-0 expression (\ref{sdc}).\footnote{For the bosonic case we may also consider a general spin-2 state. The massless case is gravity and leads to the Species scale. For a massive Spin-2 field this should match onto the spin-2 conjecture in \cite{Klaewer:2018yxi}.} 

There is a subtlety, that the relation (\ref{fwgc}) holds only when the tower contribution to the kinetic terms comes from renormalizable Yukawa couplings. As we discuss below, for the cases of the gravitino and goldstino, it turns out that also non-renormalizable couplings can lead to an order one contribution to the kinetic term through 1-loop diagrams. Therefore, the general statement is that either (\ref{fwgc}) holds, or $\chi$ has an appropriate non-renormalizable coupling of the type that goldstinos and gravitinos posses.\footnote{It is natural to expect that gravitinos and goldstinos are the only fermions which satisfy the emergence proposal through these couplings, since as we will see, they are universal and gravitational in behaviour. This suggests that any other fermions would indeed satisfy (\ref{fwgc}). Note also that a similar issue of ambiguity occurs for scalars, where again a universal coupling through dimension 5 operators suppressed by the Planck mass will satisfy the emergence proposal.} 

There are a number of key differences between (\ref{fwgc}) and gauge and scalar cases. First, note in (\ref{Yukco}) we introduced both a scalar and a fermion for each level of the tower of states, rather than just a single state at each level. This is our first encounter with a suggestion that the tower of states should be supersymmetric. Actually, we could have induced the running with just one state at each level through couplings of type $Y_n h \chi \psi_n$ or $Y_n h_n \chi \psi$. This is certainly a possibility that is consistent with the emergence proposal. However, it is strange in the sense that it does not associate the relevant states purely to the tower. In other words, it raises the question of what distinguishes the special field $h$ or $\psi$ which is not part of the tower.\footnote{For the case of $\psi$ we may consider an identification $\psi = \chi$ natural.} 
Also, as briefly mentioned in section \ref{sec:disc}, in the case where the fermion carries some charge it is natural to pair up states from different levels in the tower which would only work with something analogous to (\ref{Yukco}) (with a shifted level and charge between the fermion and the boson). Therefore, we believe that (\ref{Yukco}) is the more natural expectation. 

Another difference is that the Yukawa couplings increasing with level (\ref{Yukn}) does not arise naturally for a fermion. For a gauge field it relates to the increasing charge of states, while for a scalar field it arises because the coupling relates to the mass of the states which is increasing up the tower.\footnote{Note that the dependence on the the tower level $n$ may be modified, say as $\sqrt{n}$, but some dependence is expected. It is, however, not strictly necessary, it is possible to integrate out a tower of states with the same charge say, for example as discussed in \cite{Grimm:2018ohb}.} If the tower is supersymmetric then the increasing Yukawa coupling (\ref{Yukn}) is explained by the relation to either a gauge or scalar superpartner. So again we see that supersymmetry is naturally introduced. 

An important difference, related to the previous point, is that the overall Yukawa factor $Y$ in (\ref{Yukn}) is not normalised across all the states which couple to the fermion $\chi$. By this we mean that, say for a gauge field, charge quantization implies that the $g$ factor is universal to all the charged states. So it is not possible to make one charged state parametrically weakly coupled without making all the charged states weakly coupled.\footnote{This is true at least in a diagonal basis of gauge fields. By appropriate mixing, say through kinetic terms, it is possible to induce fractional charges.} In general, as discussed in section \ref{sec:dis}, this is no longer true for a scalar which can couple arbitrarily weakly to a state, while maintaining strong coupling to other states. For a pseudo-scalar there is an analogous normalization where charge quantization is replaced by instanton number.\footnote{This is part of the axion version of the WGC. There is also a version for higher dimension objects, which corresponds to a quantization condition for higher anti-symmetric degree tensors.} The case of a fermion is similar to a generic scalar, there is no normalization in general. Say we consider a coupling of $\chi$ to some particular field, which is not part of the tower, which is very weak. This does not imply that the coupling to the tower $Y$ must be small. So having a fermion with very small Yukawa coupling to some fields does not imply a light tower of states.\footnote{In the case where the tower is supersymmetric and the fermion is part of a vector multiplet, so a gaugino, then this normalisation is fixed again by supersymmetry.}  This makes (\ref{fwgc}) less constraining than its vector version. Note that if the small Yukawa coupling is induced by the normalisation of the kinetic term of the fermion, then such a suppression of the coupling would be universal and so include the coupling to the tower of states. Such a scenario would then be more strongly constrained by (\ref{fwgc}). 

In the case of a Spin-$\frac32$ field, a consistent microscopic description of such a fundamental field requires that it should be the gravitino. So it intrinsically requires supersymmetry, and we discuss it in the next section.

\section{Supersymmetry}
\label{sec:susy}

We have seen that there are natural suggestions that supersymmetry should be approximately restored at the tower mass scale $m_{\infty}$. In that case, all fermions are related to some field of spin 0, 1 or 2. Therefore, their Swampland constraints follow from the bosonic ones. The simplest case is when the fermion is part of a vector multiplet (so a gaugino), which then relates the overall Yukawa coupling $Y$ in (\ref{Yukn}) to the gauge coupling $Y=g$. In the case of a chiral multiplet we have the relation to the scalar coupling $Y=\mu$.

The gravitino $\Psi_{\mu}$ couples purely through operators suppressed by the Planck mass
\be
\frac{1}{M_p} \overline{\Psi}_{\mu} S^{\mu} \;.
\label{gravint}
\ee
Here $S^{\mu}$ is the supercurrent which carries dimension $\frac72$ and contains terms like 
\be
S^{\mu} \supset \left(D_{\nu} h \right)^* \gamma^{\nu} \gamma^{\mu} \chi_L \;,
\label{supcuter}
\ee
where $h$ is a boson and $\chi_L$ a (left hand component of a) fermion. See, for example, \cite{Rychkov:2007uq} for a good account of gravitinos. 

We can do a simple dimensional analysis of the 1-loop contribution from gravitino interactions to their kinetic terms. These should be suppressed by $M_p^2$ and be dimensionless and so, up to irrelevant logarithmic factors, should go as $\left(\frac{\Lambda}{M_p}\right)^2$ where $\Lambda$ is the cutoff scale of the theory. Taking $N$ fields running in the loop, and the cutoff scale as the Species scale, gives the contribution to the kinetic term
\be
N \left(\frac{\Lambda_s}{M_p}\right)^2 \sim 1 \;.
\ee
As expected from supersymmetry, this is consistent with the emergence proposal. 

Another interesting fermion is the Goldstino $\chi_G$. Upon $F$-term supersymmetry breaking the Goldstino has two interaction terms (see, for example \cite{Rychkov:2007uq})
\be
\frac{1}{F} \overline{\chi}_G \partial_{\mu} S^{\mu}_{m} + \frac{1}{M_p} \chi_G \gamma_{\mu} \overline{S}^{\mu}_m \;.
\label{goldint}
\ee
Here $S^{\mu}_m$ is the contribution to the supercurrent from the matter fields excluding the Goldstino.

It is interesting to consider the first term in (\ref{goldint}). In flat space, the F-term is related to the gravitino mass $m_{\frac32}$ as $F \sim m_{\frac32} M_p$. The divergence of the supercurrent is only non-vanishing due to supersymmetry breaking and so goes approximately as $(m_{\frac32})^2$. This yields a Yukawa coupling of the goldstino which satisfies $Y_G M_p \sim m_{\frac32}$. Comparing with (\ref{fwgc}) we would have that $m_{\infty} \sim m_{\frac32}$ which would mean that there is no effective theory of the supersymmetry breaking.  

The second term in (\ref{goldint}) allows us to avoid this conclusion. It is of a similar form as the gravitino interaction (\ref{gravint}) and so yields a contribution of order one. This is expected since the Goldstino becomes the helicity-$\frac12$ polarization of a massive gravitino.

\section{Supersymmetry breaking scale}
\label{sec:susybrsc}

In section \ref{sec:fer} we proposed a number of motivations for including supersymmetry at the tower mass scale $m_{\infty}$ in order to ensure emergence for fermions. As expected, and outlined in section \ref{sec:susy}, if supersymmetry is restored at that scale then the fermionic requirements automatically follow from the more familiar bosonic ones. 

Motivated by these ideas, we are therefore led to a natural proposal that a Swampland constraint is that supersymmetry is approximately restored at the tower mass scale for the towers associated to fermions as in (\ref{fwgc}). So, denoting the supersymmetry breaking scale $m_{SUSY}$, we have that
\be
m_{SUSY} < m^Y_{\infty} \;.
\label{msusyconj}
\ee

Actually, we will see that some refinements of the statement (\ref{msusyconj}) are required. More precisely, $m_{SUSY}$ should be the supersymmetry breaking scale sensed by the tower of states. So the mass splitting of the bosons and fermions in the tower. As well as of course the fermion $\chi$ itself. If there are multiple couplings and associated towers, then each tower has an associated $m_{SUSY}$. It may be that supersymmetry is broken up to a higher scale in some other sufficiently sequestered sector of the theory. This possibility seems to be suggested by examples in string theory. In terms of constraints on the effective theory for the fermion $\chi$ this is not so important, the effective theory essentially feels the restoration of supersymmetry below the scale in (\ref{msusyconj}).  

While the motivation for (\ref{msusyconj}) holds for fermions, it is possible to consider a stronger version of it that it should hold for any of the infinite towers. So this strong version would state that (\ref{msusyconj}) holds for any of $m^g_{\infty}$, $m^{\mu}_{\infty}$ and $m^Y_{\infty}$,  as in (\ref{mwgc}), (\ref{sdc}) and (\ref{fwgc}). Heuristically, this can also be motivated by the idea that these are ultraviolet scales where physics begins to complete towards quantum gravity, and so naturally supersymmetry is expected to make an appearance at those scales.

Such a strong version appears, at least naively, too strong. For example, we may consider a non-supersymmetric theory and compactify on a circle. The scale $m^g_{\infty}$ for the graviphoton is then the Kaluza-Klein scale. We can send this to zero by decompactifying, this would take us to the original higher dimensional theory which has a non-vanishing supersymmetry breaking scale. More generally, this example is a proposal for a setup where we have a gauge coupling which can be varied independently of the supersymmetry breaking scale. Whether something like this can be realised in string theory is not completely clear. This is because all coupling constants are dynamical and so must also be fixed by some potential which can tie them to the supersymmetry breaking effects. In section \ref{sec:sttest} we will study some simple string theory cases, and there at least it seems even the stronger version holds. On balance, we remain agnostic about the strong version. It would be very interesting to see if there is more evidence for it from string theory, or if there are good counter examples.\footnote{See, \cite{Bonnefoy:2018tcp,Bonnefoy:2020fwt} for some initial studies of the Swampland conjectures in string theory setups with broken supersymmetry.}  

%As discussed in section \ref{sec:fer}, for the general scalar and fermion cases the tower couplings $\mu$ and $Y$ need not be related to all the couplings of those fields. So (\ref{msusyconj}) is not a statement about any dimensionless coupling, but about the coupling to the tower. For the gauge case (and gauginos), charge quantization does relate the coupling to the tower to the coupling to other states, and in that sense is a much stronger statement. 

\section{Preliminary Tests in String Theory}
\label{sec:sttest}

A supersymmetric vacuum of string theory will satisfy the conjecture (\ref{fwgc}) as long as it satisfies the Swampland constraints (\ref{mwgc}) and (\ref{sdc}). As discussed in section \ref{sec:dis}, the constraint (\ref{sdc}) is actually different from the distance conjecture, and so while it has passed many tests in string theory, and we discussed some further ones in appendix \ref{app:sttest}, it is still not as well established as the original magnetic Weak Gravity Conjecture (\ref{mwgc}). Still, we will assume that it holds here. If we then break supersymmetry below the tower mass scale for the fermions, as proposed to be required by (\ref{msusyconj}), then the fermionic constraint (\ref{fwgc}) will also be satisfied since the tower will not feel the effect of supersymmetry breaking. Therefore, in searching for violation of (\ref{fwgc}) we first require a violation of (\ref{msusyconj}). 

We are not aware of any violation of (\ref{msusyconj}) in string theory. Since $Y$ is a coupling to an infinite tower of states a suppression of it would be expected to occur through parameters controlling the kinetic terms of the fields. In this section we will study such universal suppression of all Yukawa couplings through dependence on the string coupling and volume factors. 

In looking for violations of (\ref{msusyconj}) it is natural to look for settings in string theory which break supersymmetry as strongly as possible. There are a number of completely non-supersymmetric string theories \cite{AlvarezGaume:1986jb,Dixon:1986iz,Seiberg:1986by}. While all of these exhibit instabilities, we may consider them as testing grounds. As shown in \cite{Dienes:1994np} (see, for example, \cite{Abel:2015oxa,Abel:2016vex} for recent expositions), non-supersymmetric strings do not restore full supersymmetry at any scale but rather exhibit `misaligned supersymmetry'. However, the effective theory below the String scale effectively restores supersymmetry at the string scale, and it is only a tower of massive states which break it. This is the same as in Scherk-Schwarz supersymmetry breaking where there is a tower of winding modes which exhibit a non-supersymmetric spectrum even if the size of the extra dimension is taken to infinity. The low-energy theory however does exhibit supersymmetry in that limit, with only exponentially small corrections coming from the massive winding modes. Further, even in misaligned supersymmetry the mass splitting of the states is always of order their mass, and so is effectively satisfying (\ref{msusyconj}) at all levels with the tower mass scale being the string scale. All in all, it seems that effectively supersymmetry is always restored at the string scale.

\subsection{String coupling dependence}

As an initial test, we may therefore consider if $m^Y_{\infty}$ can be made parametrically smaller than the string scale. We will analyse this generally with the parameter being the string coupling $g_s$. We keep track of the factors of string coupling as follows. We absorb factors of $g_s$ in the vertex operators into the fields. The factor of $g_s$ for a given operator in the theory is therefore determined by the topology of the amplitude which determines it. We will only be concerned with tree-level operators and so they come with a prefactor of $g_s^{-p}$ with $p=1$ for open strings (disc) and $p=2$ for closed strings (sphere). We then take the fields to be dimensionless and complete the dimensions with the only mass scale in the theory, the string scale $M_s$. So for a $D$-dimensional effective theory the kinetic terms for scalar $\phi$, fermion $\chi$ and their Yukawa coupling take the schematic form
\be
\frac{1}{g_s^p} M_s^{(D-2)} \left( \partial \phi \right)^2 + \frac{1}{g_s^p} M_s^{(D-1)} \chi \partial \chi + \frac{1}{g_s^p} \hat{Y} M_s^{D} \phi \chi \chi \;.  
\label{mskiny}
\ee 
In $D$-dimensions the relation (with respect to the string coupling dependence) between the string scale and Planck mass is
\be
M_s \sim g_s^{\frac{2}{D-2}} M_p \;.
\label{Msrel}
\ee
Now we need to canonically normalise the fields by absorbing appropriate powers of $g_s$ from the kinetic terms. So write the cubic coupling as
\be
Y \tilde{\phi} \tilde{\chi} \tilde{\chi} M_p^D \;,
\ee
with
\be
\tilde{\phi} = g_s^{1-\frac{p}{2}} \phi \;,\;\; \tilde{\chi} = g_s^{\frac{D-1}{D-2}- \frac{p}{2}} \chi \;,
\ee
the (dimensionless) canonically normalised fields.
The canonically normalised Yukawa coupling $Y$ then has $g_s$ dependence of 
\be
Y \sim \hat{Y} g_s^{\frac{\left(\frac{p}{2}-1\right)D+\left(4-p\right)}{D-2} } \;.
\label{Yhatrel}
\ee
Comparing (\ref{Msrel}) and (\ref{Yhatrel}) we find that taking $g_s \rightarrow 0$ contributes towards ensuring $m^Y_{\infty}>M_s$ whenever $D>2$. For closed strings $p=2$ we have that $m^Y_{\infty} \sim M_s$ for any $D$, while for open strings $p=1$, $m^Y_{\infty}$ is parametrically higher than $M_s$ at weak coupling. 

We therefore conclude that the string coupling $g_s$ cannot be used to violate (\ref{msusyconj}), and for open-string fields always works towards satisfying (\ref{msusyconj}) more strongly. In particular, this means that in non-supersymmetric string theories, where the only parameter is $g_s$, (\ref{msusyconj}) is satisfied.

Actually, even the stronger version where the susy breaking scale is smaller than the towers associated to any couplings is satisfied. So say for the $SO(16)\times SO(16)$ Heterotic string we have $m^g_{\infty} \sim M_s$. 

\subsection{Volume dependence}

A way to break supersymmetry through dimensional reduction which is compatible with string theory is through the Scherk-Schwarz mechanism \cite{Scherk:1978ta}. More generally, this can be understood as compactifying on manifolds with torsion.\footnote{By duality, this should also capture supersymmetry breaking by fluxes.} Let us consider such a setup with a compactification average scale $R$. The scale of supersymmetry breaking is then the Kaluza-Klein scale $M_{KK}$. Therefore a parametric test of (\ref{msusyconj}) is whether it is satisfied with $m_{SUSY} \sim M_{KK}$, for any value of $R$. 

To illustrate the point we will ignore any dependence on the string coupling, and consider dimensional reduction on a circle (the case of reduction on higher dimensional manifolds will follow simply). Let $\hat{M}_p$ denote the higher dimensional Planck mass, say in $D$ dimensions. Then we follow a similar analysis as in (\ref{mskiny}), so start with  
\be
\hat{M}_p^{(D-2)} \left( \partial \phi \right)^2 + \hat{M}_p^{(D-1)} \chi \partial \chi +  \hat{Y} \hat{M}_p^{D} \phi \chi \chi \;,
\label{mpkiny}
\ee 
with $\hat{Y}$ the Yukawa coupling in the higher dimensional theory. 
%We are only interested in inducing small Yukawa couplings through the volume dependence of the compactification, so we can henceforth set the higher dimensional Yukawa to one $\hat{Y} \sim 1$.

After the dimensional reduction the lower dimensional Planck mass $M_p$ is related to the higher dimensional one through
\be
\frac{\hat{M}_p}{M_p} \sim \frac{1}{\left(RM_p \right)^{\frac{1}{D-2}}} \;.
\ee
Using this we find that $m_{\infty}^Y$, associated to the canonically normalised lower dimensional Yukawa coupling $Y$, takes the form
\be
m_{\infty}^Y \sim Y M_p \sim \hat{Y} \hat{M}_p \;.
\ee
The Kaluza-Klein scale is
\be
M_{KK} \sim \frac{\hat{M}_p}{RM_p} \sim \frac{m_{\infty}^Y}{\hat{Y}  R M_p} \;. 
\ee
We therefore see that the $R$ dependence works towards satisfying (\ref{msusyconj}). This analysis is simple to generalise for higher dimensional reductions and therefore suggests that when breaking supersymmetry by compactification effects, increasing the volume of the compactification will always work towards respecting (\ref{msusyconj}).

It is worth noting that also in Scherk-Schwarz breaking of supersymmetry, the stronger version of (\ref{msusyconj}) applied also to the gauge coupling holds, with the gauge field here being the graviphoton that has an associated $m^g_{\infty} \sim M_{KK} \sim m_{SUSY}$. 

The analysis of the volume scaling is relevant for Yukawa couplings of bulk fields which are not localised in the extra dimensions. If we consider fields localised, for example on D-branes, then the expectation is that their interaction strengths will only get stronger since their wavefunctions are less diluted. If supersymmetry breaking is mediated gravitationally to their sector, and therefore is diluted by the full volume, then (\ref{msusyconj}) should continue to hold. 

We can also consider complicated scenarios of supersymmetry breaking in string theory (which are under less microscopic control). For example, the Large Volume Scenario \cite{Balasubramanian:2005zx} proposes a non-supersymmetric vacuum of type IIB string theory. In this proposal the supersymmetry breaking scale is controlled by the volume ${\cal V}$ as $m_{SUSY} \sim {\cal V}^{-1}$. Yukawa couplings for fields living on D-branes behave as $Y \sim {\cal V}^0$ and $Y \sim {\cal V}^{-\frac12}$ depending on if the branes are on contractible or large cycles \cite{Conlon:2006tj}. While these are Yukawa couplings between massless fields, they can be taken as an indication for the coupling to the massive tower, and so suggest that (\ref{msusyconj}) is satisfied. 

We may consider a strongly warped region or throat in the extra dimensions and attempt to suppress the Yukawa couplings this way. However, the towers of states, say those tied to the string scale or Kaluza-Klein scale, are also reduced due to the warping. Or in other words, the system inside a warped throat contains not just the fermion but also the full tower of states, and so their coupling is not expected to be suppressed by this. It would be interesting to perform a more quantitative analysis of this. 

Finally, we can consider how the strong version of (\ref{msusyconj}), where we may replace the tower mass scale by $m^g_{\infty}$ say, fairs in string examples. For a specific example, we may take the Large Volume Scenario above, where we can make the gauge coupling of D7 branes very small. These wrap four-cycles and so we should have $g \sim {\cal V}^{-\frac13}$, which satisfies (\ref{msusyconj}). There is also a general interesting relation for closed-string couplings (which may be gauge or higher form). These, by the Weak Gravity Conjecture, bound the tension of associated D-branes. If we then break supersymmetry by introducing non-supersymmetric D-branes, then the supersymmetry breaking scale should be at most of order the brane tension and so be bounded from above by the associated coupling.

\section{Discussion}
\label{sec:disc}

In this note we considered what Swampland conditions may be associated to the presence of fermions in the theory. Motivated by the idea of emergence, we proposed that there must exist an infinite tower of states which couple to any fermion in the theory. The mass scale of this tower is set by the strength of the Yukawa coupling of the fermion to the tower (\ref{fwgc}). This is a fermionic version of the (magnetic) Weak Gravity Conjecture.  

While the fermionic constraint is analogous in many ways to the vector field Weak Gravity Conjecture, there are some fundamental differences. Perhaps foremost is the point that for a generic fermion (and also a generic scalar) there is no overall normalization of the Yukawa couplings, which means it can couple very weakly to some particle while still coupling strongly to the massive tower of states. This implies that the conjecture does not necessarily imply a light tower of states if a fermion has any small Yukawa coupling to any field. In this sense, it is unfortunately less predictive. However, still, a fermion cannot couple weakly to all fields without having a light tower. In particular, this restricts the kinetic terms of the fermion, which affect all couplings.

The fermionic constraint appeared to be tied to the presence of supersymmetry at the tower mass scale. This motivated a further proposal that for the tower of states associated to fermions, supersymmetry must be effectively restored by the mass scale of the tower (\ref{msusyconj}). So if a fermion couples weakly to all fields, then supersymmetry must be broken at a low scale.

We made some preliminary checks of these proposals in string theory, and these were consistent with them. 

As a precursor to the fermionic analysis we also, in section \ref{sec:dis} and appendix \ref{app:sttest}, clarified some aspects of the constraint (\ref{sdc}). In particular, emphasising its independence from the distance conjecture, and performing some simple tests in string theory of its validity.  

An aspect which we did not explore in detail is the role fermion charges under symmetries play in the conjecture. It would be interesting to explore how such a charge would effect the conjecture and what this could imply for the charges of the towers of states. In particular, it is natural to expect that if the fermion is charged then the tower should have increasing charges. This way one is able to always form a neutral Yukawa coupling by taking states from different levels in the tower. A small piece of evidence towards this is that a way to make open-string modes couple weakly is by diluting their wavefunction. This means making the cycle wrapped by the brane they are on very large. This is the same limit as making the gauge coupling small, which is proposed to lead to a light tower of charged states of increasing charges. It would be interesting to perform more quantitative tests of this. 

\vspace{10px}
{\bf Acknowledgements}
%\vspace{10px}
\noindent
I would like to thank Gia Dvali, Dieter L\"ust, Irene Valenzuela and Timo Weigand for useful discussions.

\appendix

\section{Scalar couplings in string theory}
\label{app:sttest}

In this appendix we make some comments on, and very simple tests of, (\ref{sdc}) in string theory for cases where $\mu$ is not controlled by the scalar expectation value.\footnote{There was an analysis in \cite{Blumenhagen:2019qcg} of the emergence proposal at finite distances in field space which is in the same spirit. We will consider much simpler settings in this section.} The case where $\mu \rightarrow 0$ coincides with $\phi \rightarrow \infty$ corresponds to the Swampland Distance Conjecture which has been extensively tested.

One setting where general checks can be made is for scalars controlling the volumes of cycles, and for their axionic superpartners. For concreteness we can consider the Kahler moduli of type IIA compactifications on a Calabi-Yau to four dimensions, labelled here as $\phi$, but the argument is, through dualities, general. The Kahler moduli have associated towers of states corresponding to wrapped (even dimensional) D-branes. The mass scale of this tower schematically behaves as (see, for example \cite{Palti:2019pca})
\be
m_{\infty} \sim \left( \frac{M_s}{M_p} \right) \phi\;.
\label{mcl}
\ee
This can be understood by noting that $\phi$ gives the volume of the cycle wrapped by the brane, and the mass should go like the string scale. Now we consider the coupling of the fields to the branes, 
\be
\mu \sim \partial_{\phi} m_{\infty}  \sim \frac{M_s}{M_p} \;.
\label{mucl}
\ee
Actually, (\ref{mucl}) is not quite right for two reasons. First, since the volume of the cycle $\phi$ is not a canonically normalised field (the canonically normalised field actually leads back to exactly (\ref{mcl}) again in examples we could study). But this is not important because of the limit we will consider. The second reason is that $M_s$ depends on $\phi$, but taking a derivative of it only induces a subleading additional term in (\ref{mucl}). 

The point is then rather simple, we can send $\mu \rightarrow 0$ by varying the other fields such that $\frac{M_s}{M_p} \rightarrow 0$. But clearly $m_{\infty}$ would retain the same parametric dependence. In such a setting, the scalar $\phi$ has a fixed expectation value for the $\mu \rightarrow 0$ limit, but still (\ref{sdc}) is satisfied. In fact, we could apply the same logic to the axion superpartner of $\phi$, and the result would be the same, this time holding for the case where the scalar has vanishing expectation value. While this analysis seems simple and obvious, it actually underlies (through dualities) the general $\mu \rightarrow 0$ limits in field space of closed string moduli, as studied for example in \cite{Grimm:2018ohb,Grimm:2018cpv,Corvilain:2018lgw}.

The previous analysis is general for towers of wrapped states, but it is important to allow for the possibility that those states are not a tower of particles but possibly to some extended objects (which exhibit a tower of states). For example, strings with a tower of oscillator modes (for example, as in \cite{Lee:2018spm,Font:2019cxq,Lee:2019wij,Marchesano:2019ifh,Baume:2019sry}). The mass scale $m_{\infty}$ would then be associated to the tension of the objects. As discussed in many places, see for example the review \cite{Palti:2019pca}, allowing for such a generalisation is important in the case of scalar fields (and will also be true for fermions). 

More evidence is that in the case of ${\cal N}=2$ supersymmetry one can prove that for any scalar field in the vector multiplet moduli space there exists an infinite number of charges whose associated BPS states satisfy $\mu M_p > m_{\infty}$ \cite{Palti:2017elp}. This proof holds at any point in moduli space, so is unrelated to the expectation value of the particular scalar which couples to the BPS states. In all known points in moduli space this inequality is an approximate equality, thereby satisfying (\ref{sdc}).

Informative tests would be in the context of open-string fields, especially since they are charged under gauge symmetries. However, in general the tower of states associated to open string modes is difficult to analyse quantitatively as it is expected to be composed of non-perturbative states \cite{Lee:2018spm}. Nonetheless, since open string fields can be uplifted to geometry in F-theory and  M-theory, it is natural to expect that they will also satisfy the constraint. It would be good to test this in future work.

\bibliography{Higuchi}

\end{document}